\documentclass{article}

\usepackage{arxiv}

\usepackage[utf8]{inputenc} 
\usepackage[T1]{fontenc}    
\usepackage{hyperref}       
\usepackage{url}            
\usepackage{booktabs}       
\usepackage{amsfonts}       
\usepackage{nicefrac}       
\usepackage{microtype}      
\usepackage{lipsum}
\usepackage{graphicx}
\usepackage[authoryear]{natbib}
\usepackage{amsmath}
\graphicspath{ {./images/} }

\title{From Winter Storm Thermodynamics to Wind Gust Extremes: Discovering Interpretable Equations from Data}

\author{
Frederick Iat-Hin Tam,
Fabien Augsburger,
Tom Beucler \\
Faculty of Geosciences and Environment \\
University of Lausanne \\
Lausanne, VD, Switzerland
}

\begin{document}
\maketitle
\begin{abstract}
Reliably identifying and understanding temporal precursors to extreme wind gusts is crucial for early warning and mitigation. This study proposes a simple data-driven approach to extract key predictors from a dataset of historical extreme European winter windstorms and derive simple equations linking these precursors to extreme gusts over land. A major challenge is the limited training data for extreme events, increasing the risk of model overfitting. Testing various mitigation strategies, we find that combining dimensionality reduction, careful cross-validation, feature selection, and a nonlinear transformation of maximum wind gusts informed by Generalized Extreme Value distributions successfully reduces overfitting. These measures yield interpretable equations that generalize across regions while maintaining satisfactory predictive skill. The discovered equations reveal the association between a steady drying low-troposphere before landfall and wind gust intensity in Northwestern Europe. 
\end{abstract}


\section{Introduction}
Damaging windstorms associated with extratropical cyclones are among the leading weather-related disasters in the mid-latitudes winters, especially in Europe \citep{schwierz2010}. From a weather forecasting standpoint, early detection of reliable precursors leading to extreme windstorms helps timely disaster mitigation. Beyond horizontal temperature differences strengthening storms \citep{Laurila2021}, studies have highlighted how lower-atmosphere humidity fuels wind jets \citep{Martínez‐Alvarado2013} and transports high-wind air toward the surface \citep{browning2015}, yet simple models linking winterstorm characteristics to wind gusts are lacking.  

Interpretable machine learning has enabled trustworthy data-driven models of thunderstorms \citep{Hilburn2023} and tropical cyclones \citep{tam2024}, improving process understanding by reducing the attribution uncertainty of post-hoc explainable AI methods \citep{Mamalakis2023}. This motivates us to apply data-driven methods to uncover novel temporal patterns in environmental precursors to extreme European windstorms. A key challenge is the limited availability of extreme storm data, which restricts the use of complex nonlinear models prone to overfitting in small-sample conditions. However, this limitation also presents an opportunity to derive a simple, interpretable equation approximating complex physical processes, paving the way for data-driven discovery. 

Data-driven equation discovery distills simple laws from empirical data \citep{schmidtlipson2009} with methods ranging from sparse selection algorithms to symbolic regression \citep{song2024towards}. While data-driven approaches have successfully identified physical laws from controlled fluid dynamics experiments data \citep{brunton2016,ChengAlkhalifah2024}, finding interpretable equations from noisy, multidimensional weather and climate data remains under-explored. Balancing model performance and algorithmic complexity, \citep{Grundner2024} uncovered simple analytical cloud cover equations from high-resolution model outputs. We similarly implement a hierarchical modeling framework to track the trade-off between the model complexity and generalization capability, identifying a set of sparse yet descriptive equations that achieves satisfactory prediction skills for unseen storm cases across European regions. Additionally, we explore the best strategy for dimensionality reduction and feature selection for reliable discovery of simple equations from small yet high-dimensional datasets.

\section{Data}
\label{sec:Data}
\subsection{Raw Datasets Description}
\paragraph{Winter Windstorm Tracks for Europe}
\label{winter_dataset}
The winter storm data analyzed in this study is based on the ECMWF Winter Windstorm Indicator dataset \citep{winterERA5}. This dataset documents historical severe windstorms between 1979 and 2021, and their footprints. The tracking of storms is performed by applying a tracking algorithm on the ERA5 reanalysis data from October through March. The tracking algorithm tracks local maxima of 850 hPa relative vorticity, with a minimal 10m wind speed threshold of 25 m s$^{-1}$. A total of 118 storms is identified in the 42-year period. This study analyzes storms in the last 30 years (96 storms). Since we focus on the impact of severe windstorms over land in this study, short-lived or non-landfalling storms are filtered out, which leaves us with 63 storms.

\paragraph{ERA5 Meteorological Reanalysis} \label{data:era5}
We use the hourly ERA5 reanalysis dataset with a horizontal resolution of 0.25$^o$-0.25$^o$ to characterize the environmental conditions for all tracked storms. Our analysis considers 28 potential environmental drivers for wind gust formation and intensity \citep{schulzlerch2016} listed in Table~SI6. To find the driver conditions local to each storm, we adopt a system-following 4$^o$-4$^o$ grid box centered around the storm track of all 28 environmental drivers. These drivers include measurements of the winterstorm kinematics (e.g., 10m wind speeds), precipitation characteristics (e.g., mean precipitation rates), and the thermodynamic characteristics (e.g., latent heat flux; K-index) near the windstorms. A full list of the ERA5 variables and their units is provided in Table S6.

\subsection{Data Preprocessing: Dimensionality Reduction and Partitioning} \label{data:dims_reduce}
Focusing on extreme events limits the training data for data-driven models. To ensure reliable equation discovery, we reduce the dimensionality of storm history and post-landfall gust data, turning our dataset into one that can be analyzed with simple data-driven models without overfitting.

\paragraph{Features: Prescribed Spatial Dimensionality Reduction} \label{data:spatialfilter}
To encode the 4D storm history fields as time series, we compute four spatial statistics: maximum (max), minimum (min), mean, and standard deviation (std). Using predefined statistics instead of learned filters facilitates physical interpretability. For example, a higher temperature standard deviation may indicate the presence of weather fronts \citep{lagerquist2020}. To further reduce dimensionality, we apply Principal Component Analysis (PCA) to compress time series into orthogonal temporal modes ($\Pi_{X} (t)$) and PC loadings ($\mathrm{PC}_{X}$), which are scalars denoting the projection of time series onto $\Pi_{X} (t)$:
\begin{equation} \label{define:PC}
    X(t)-\overline{X(t)} = \sum_{i=1}^{N} \mathrm{PC}_{X,i} \Pi_{X,i}(t), 
    \quad \text{where} \quad \left\langle \Pi_{X,i}(t) | X(t) \right\rangle_X = \mathrm{PC}_{X,i}.
\end{equation}
Here, \(N\) is the number of retained PC modes indexed by \(i\), and $\langle \cdot | \cdot \rangle_X$ denotes the inner product associated with the PCA decomposition of \(X\). The modes \(\Pi_{X,i}(t)\) are normalized such that $\left\langle \Pi_{X,i}(t) | \Pi_{X,j}(t) \right\rangle_X = \delta_{ij} $ where \(\delta_{ij}\) is the Kronecker delta, ensuring orthonormality of the PC modes.

\paragraph{Targets: Geographical Clustering} \label{data:geos_kmeans}  
In a low-sample-size regime, predicting wind gusts for every pixel across Europe is impractical and difficult to interpret. We use K-means++ clustering \citep{arthur2007kmeans} to partition the landmass into \(K\) subregions, incorporating elevation, latitude, longitude, and maximum gusts from 63 severe windstorms. This ensures clusters reflect windstorm gust characteristics while remaining spatially coherent. To enforce geographical contiguity, we apply k-nearest neighbors \citep{enaChoi1986}, reassigning edge pixels based on the most frequent cluster label. Although silhouette scores suggest an optimal \(K = 4\)–5 (Fig.~S1), these clusters group disjointed regions with distinct wind gust characteristics (Fig.~S2). We therefore consider three additional clustering metrics (e.g., \citep{XieBeni1991}), which identify local optima at \(K = 9\), 14, and 15 (Fig.~S3). Among these, \(K = 15\) best preserves geographical compactness and aligns with established European climatic zones \citep{Jylha2010} (Fig.~S4). With clusters defined, we extract, for each storm, the highest instantaneous 10m wind gust within each cluster during the 15 hours post-landfall. This window ensures short-lived but intense storms like Lothar \citep{wernli_etal2002} are included. Our target is the maximum 15-hour wind gust within each cluster (\(U_{\text{gust}, i}\)), where \(i\) indexes the 15 clusters.

\paragraph{Cross-Validation Procedure} \label{data:cross_val}  
To ensure data-driven models select gust predictors applicable to severe windstorms, we always leave out the same 7 storms for testing. The test set is randomly chosen, except for the Lothar storm, which is explicitly withheld to assess performance on an extreme, unseen case. The remaining 56 storms are split into 7 training-validation folds by randomly selecting 7 sets of 8 storms without replacement for validation ("Random split"). Alternatively, as a step toward invariant causal prediction \citep{Peters2016}, we identify features robust to distributional shifts by clustering the output vector \(\boldsymbol{U_{\mathrm{gust}}}\) into 7 groups via K-means++ (Table S2). We then train on 6 clusters and validate on the remaining one, repeating this process 7 times (``ICP split''), following \citep{Hafner2023}. \citep{sweet2023cross} shows that such an ``ICP split'' outperforms naive k-fold cross-validation for regression. 

\subsection{Extreme Value Theory for Wind Gust Distributions} \label{text:EVT}  

\begin{figure}[htbp]
  \centering
  \includegraphics[width=0.99\textwidth]{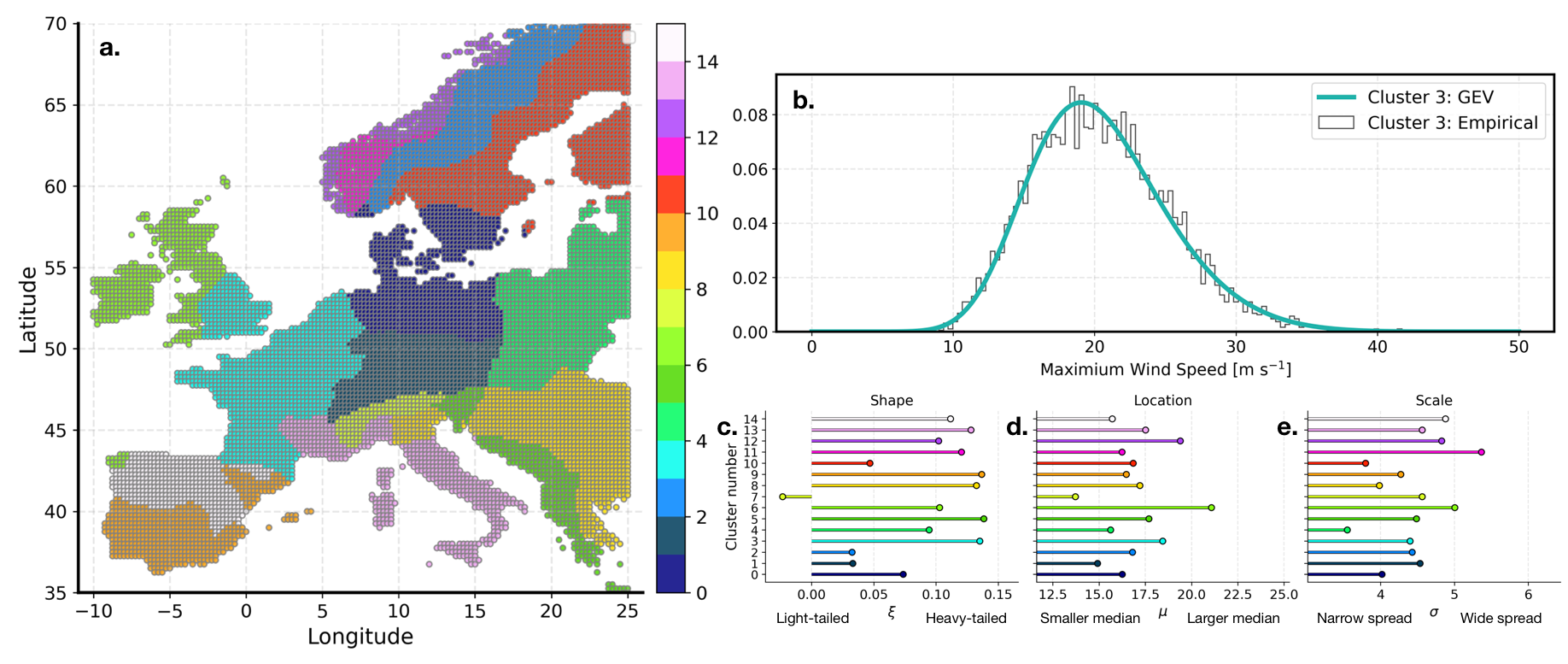}
  \caption{(a) We derive geographically contiguous regions of the European continent via K-means++ and kNN clustering to define maximum wind gust targets. (b) Extreme gust distributions are modeled using the GEV distribution, with shape (c), location (d), and scale (e) parameters fitted for each region.}
  \label{fig:clustermaps}
\end{figure}

Assessing wind gust severity relative to local climatology and distinguishing moderate from extreme gusts are key to risk assessment. We describe extreme gusts using the Generalized Extreme Value (GEV) distribution, a family of continuous distributions designed for block maxima such as extreme sea levels \citep{mendez2007}. The cumulative distribution function (CDF) and its functional inverse (quantile function) are  

\begin{equation} \label{eq:GEV}
    G_{\mu,\sigma,\xi} (x) = 
    \begin{cases} 
        \exp \left[ -\left(1+\xi \frac{x-\mu}{\sigma}\right)^{-\frac{1}{\xi}}\right], & \xi \neq 0, \\
        \exp \left[ -\exp\left(- \frac{x-\mu}{\sigma}\right)\right], & \xi = 0,
    \end{cases}
    \quad G_{\mu,\sigma,\xi}^{-1} (p) =
    \begin{cases} 
        \mu + \frac{\sigma}{\xi} \left[ \left(-\ln p\right)^{-\xi} -1 \right], & \xi \neq 0, \\
        \mu - \sigma \ln (-\ln p), & \xi = 0.
    \end{cases}
\end{equation}
where \(\mu \in \mathbb{R}\) (location), \(\sigma \in \mathbb{R}_{+}^{*}\) (scale), and \(\xi \in \mathbb{R}\) (shape). Unlike empirical CDFs, the GEV enables extrapolation beyond observed data to estimate return periods for extreme events. Due to the high spatial variability in land surface and topography across Europe, we fit a separate GEV for each cluster (Fig. \ref{fig:clustermaps}c-e; Table S1). After testing multiple definitions and durations for block maxima sampling, we opted for \textit{daily} maximum gusts over all cluster grid points during winter months (Fig.~S5). This approach differentiates regions frequently impacted by strong gusts from those less affected (Fig.~\ref{fig:clustermaps}a,d) and allows modeling of unseen extremes, which empirical distributions cannot capture (Fig.~\ref{fig:clustermaps}b). 

\section{Methodology}
\paragraph{Feature Selection}  
We aim to derive \textit{interpretable} equations linking storm characteristics to post-landfall gusts. Even with prescribed spatial statistics and PCA for dimensionality reduction, the number of features (28 predictors \(\times\) 4 spatial statistics \(\times\) 1–12 PC loadings) still far exceeds the number of training and validation samples (56). To address this underdetermined problem, we apply sequential forward feature selection, minimizing the mean or maximum validation root-mean-squared error (RMSE) across our 7 cross-validation splits. Features are added until the RMSE objective no longer decreases. 

\paragraph{Model Hierarchy} \label{data:hierarchy}  
We construct a hierarchy of data-driven models expressible as simple equations by training multiple linear regressions (MLR) with PC loadings as features while varying three hyperparameters: feature smoothing, retained feature variance, and target transformation. For feature smoothing, we apply a low-pass rectangular filter to the PC loading time series, testing cutoff frequencies from 0.52 rad/hr to 1.92 rad/hr. For retained variance, we set four thresholds to determine the number of retained PC components per predictor: [75\%, 80\%, 85\%, 90\%]. The number of retained PC components per predictor acts as a regularization hyperparameter controlling the predictor variance fed to the ML framework. Finally, we explore a nonlinear, data-adaptive transformation of maximum wind gusts to encourage out-of-distribution generalization for extremes \citep{buritica2024}. For each sample, we first obtain its percentile value from the CDF fitted in each region, then apply a nonlinear transformation to better distinguish extreme values in the GEV distribution tails:  $Z_i = -\ln\left[1 - \text{CDF}_i\left(U_{\text{gust,i}}\right)\right]$, where \( Z \) is the transformed target, and \(\text{CDF}_i\) is the cumulative distribution function of the GEV fitted in region \( i \). All MLRs are trained using the least squares algorithms from the Scikit-Learn Python package \citep{pedergosa2011}. 

\paragraph{Pareto Optimality for Equation Discovery} \label{method:Pareto}  
Balancing performance and complexity is a multi-objective optimization task that can be addressed through Pareto optimization \citep{jinsendhoff2008}. Following sequential feature selection, we select optimal models from the empirical ``Pareto front'', corresponding to the lowest error achieved for a given complexity before the mean (or max) validation RMSE across 7 cross-validation splits plateaus. We define complexity as the number of unique features (predictor + spatial statistic) selected, as this better reflects equation simplicity than, for instance, the number of trainable parameters.

\paragraph{Model Evaluation}  
Pareto optimization seeks models with the lowest error while maintaining simplicity. We evaluate model performance using the mean validation RMSE across 7 folds and measure complexity by the number of selected unique features (predictor + spatial statistic). Using the max validation RMSE yields different Pareto-optimal models, which are also evaluated. For models predicting \(\boldsymbol{Z}\), we convert predictions back to wind gust speed in physical units using the per-cluster GEV quantile function (\(G_{\mu,\sigma,\xi}^{-1}\)) before computing RMSE, which ensures consistency across model hierarchies.

\section{Results}  
\subsection{Pareto-Optimal Models for Wind Gust Prediction} \label{section:Paretoplots}  

\begin{figure}%
   \centering
  \includegraphics[width=0.99\textwidth]{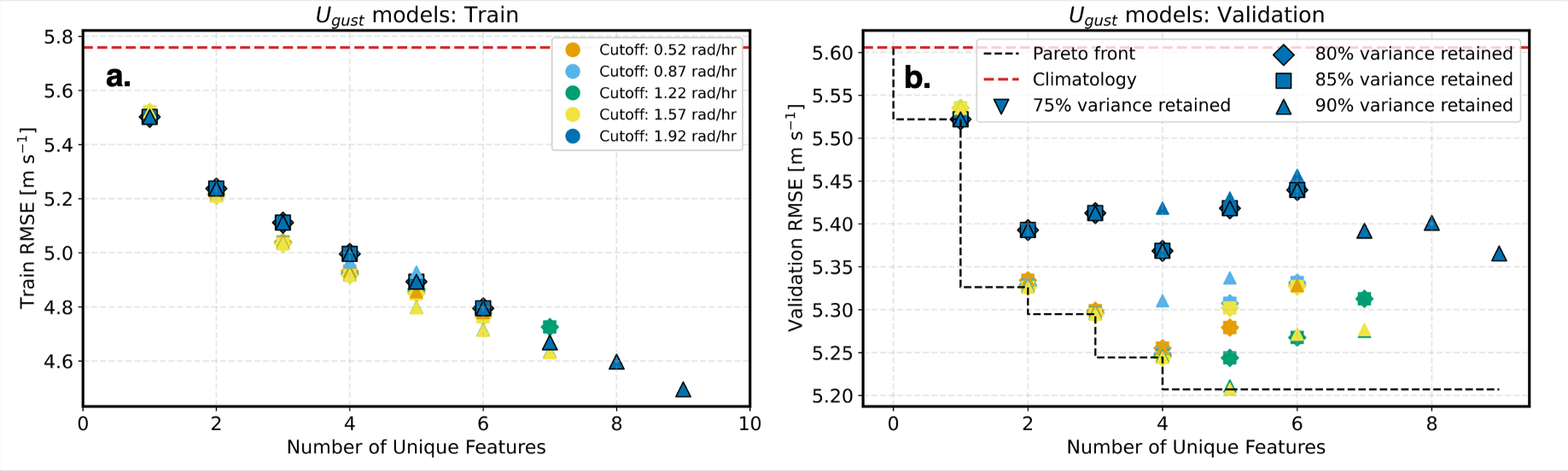}
 \caption{(a) Training and (b) validation (complexity, error) plots for MLR models predicting \( U_{\mathrm{gust}} \), with features selected to minimize the maximum validation RMSE across 7 randomly split cross-validation folds. Each colored symbol represents a model with different hyperparameters: cutoff frequency for smoothing is indicated by transparency levels, and retained feature variance by shape and color. The Pareto front (black dashed line in b) shows improved validation skills up to 5 unique features, suggesting that sparser linear models generalize better. In contrast, training RMSE decreases linearly with additional features, suggesting overfitting beyond 5 features. All models outperform the climatological baseline, which predicts the training mean \( U_{\mathrm{gust}} \) and is shown as a red dashed line}  
 \label{fig:pareto}
 \end{figure}

We explore the (complexity, error) space of trained MLRs by varying hyperparameters---feature smoothing, retained variance, and target transformation---as well as the cross-validation strategy (mean or max validation RMSE). This allows us to identify Pareto-optimal maximum wind gust equations in low sample size conditions, forming the empirical \textit{Pareto front} (dashed black line, Fig. ~\ref{fig:pareto}b). When features are selected by minimizing the max validation RMSE, the minimum achievable RMSE stabilizes after 5 selected unique variables, suggesting that additional variables may not improve generalization to unseen cases. The steady decline in validation RMSE beyond 5 features sets an upper complexity limit, beyond which models lose their ability to generalize across regions and storms with different gust spatial patterns. Additionally, the Pareto front (Fig. ~\ref{fig:pareto}b) indicates that models perform best when retaining 90\% of PC variance and applying moderate smoothing (removing oscillations with frequency \( > 1.5 \) rad/hr). These hyperparameters yield features most likely to generalize well across all storm cases.

As expected, the performance table for the best \textit{sparse} models from different feature selection methods (Table~\ref{table:Performance}) shows that models targeting the transformed gusts ($\boldsymbol{Z}$) generally underperform models directly targeting $\boldsymbol{U_{\text{gust}}}$ on the validation set when RMSE is measured in $\boldsymbol{U_{\text{gust}}}$ units. Additionally, cross-validation using the ``ICP split'' yields no validation RMSE improvement over naive random splitting. However, generalizable equations must perform robustly on unseen storms. While the ``ICP split'' sacrifices some training and validation skill, it successfully identifies sparse equations that generalize well to extreme cases like Storm Lothar, as evidenced by its low test RMSE. In contrast, equations discovered using random splits consistently perform worse on test sets, suggesting implicit overfitting to validation sets by selecting variables optimized for specific gust spatial distributions. The smaller error spread of $\boldsymbol{U_{\text{gust}}}$ models confirms their tendency to select sparse equations suboptimal for unseen cases, regardless of hyperparameters. Thus, splitting storms by their wind gust spatial distribution for cross-validation (``ICP split''), transforming the target using extreme value theory, and leveraging this information for feature selection leads to more reliable models, revealing features that better anticipate unseen cases.

\begin{table}[]
\caption{RMSE for the best models in the hierarchy, along with their unique feature count. We report the mean and the standard deviation (in parentheses) across feature selection methods and use bold font for the best RMSE. Model performance is compared to the climatology mean baseline to assess skill. }
\begin{tabular}{l|llll}
Feature selection methods     & Feature count & Training    & Validation  & Test        \\ \hline
Max RMSE; Random split; U     & 4              & 4.92 ($\boldsymbol{0.06}$) & 5.24 ($\boldsymbol{0.30}$) & 6.57 (0.02) \\
Mean RMSE; Random split; U    & 4              & $\boldsymbol{4.86}$ (0.13) & $\boldsymbol{5.16}$ (0.75) & 7.42 (0.01) \\
Max RMSE; Random split; Z     & 4              & 5.22 (0.09) & 5.62 (0.42) & 7.17 (0.02) \\
Mean RMSE; Random split; Z    & 4              & 5.04 (0.12) & 5.39 (0.61) & 7.17 (0.05) \\
Mean RMSE; ICP split; U   & 4              & 5.05 (0.14) & 5.76 (1.14) & 6.20 ($\boldsymbol{0.01}$) \\
Max RMSE; ICP split; U    & 3              & 5.20 (0.13) & 5.86 (0.93) & 6.68 (0.01) \\
Mean RMSE; ICP split; Z   & 4              & 5.12 (0.14) & 5.90 (1.15) & $\boldsymbol{6.08}$ (0.03) \\
Max RMSE; ICP split; Z    & 3              & 5.40 (0.18) & 6.00 (0.79) & 6.47 (0.02) \\ \hline
Climatology; U; Random split  &  0              & 5.75        & 5.61        & 6.23        \\
Climatology; U; ICP split &  0              & 5.76        & 5.71        & 6.23       
\end{tabular}
\label{table:Performance}
\end{table}
\vspace{-10pt}

\subsection{Reducing Geographical Bias through GEV-Informed Wind Gust Transformations} \label{section:EVTimplication}  

\begin{figure}%
 \centering
 \includegraphics[width=0.99\textwidth]{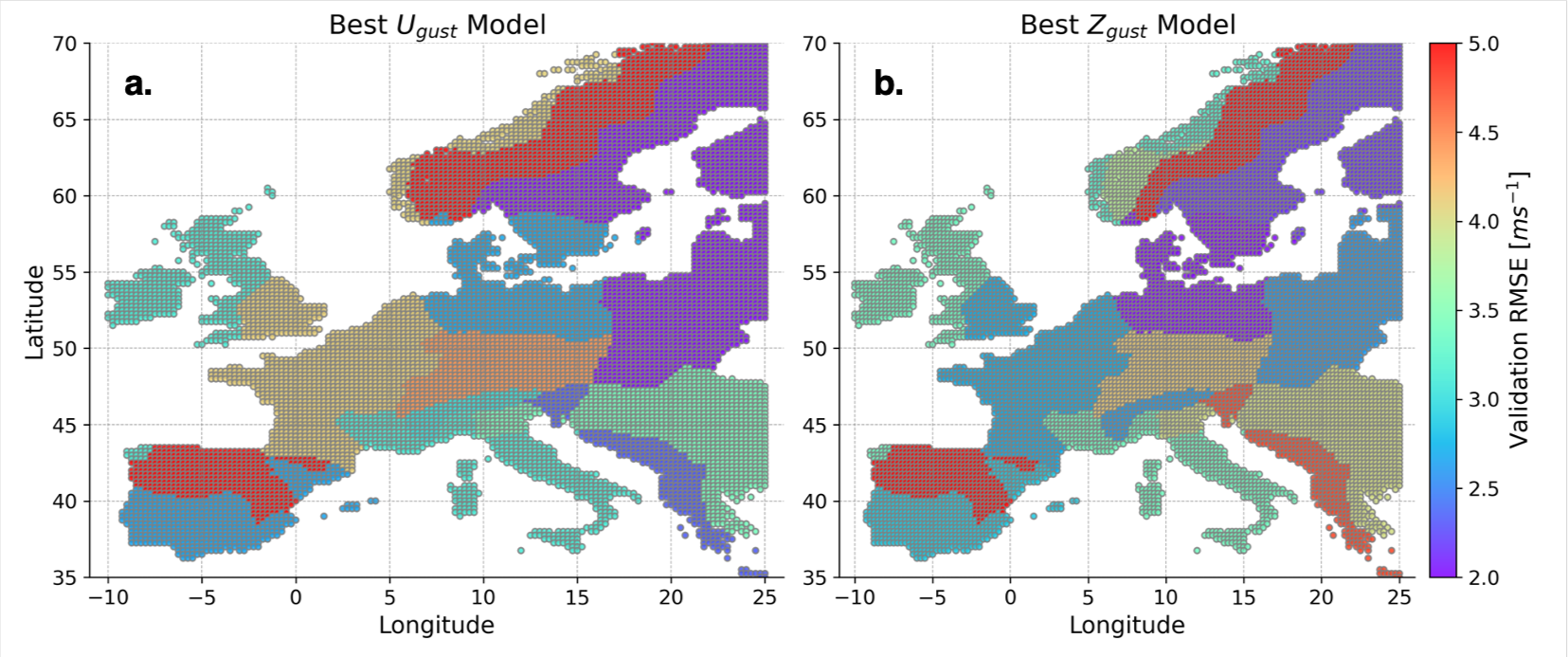}
 \caption{Validation error maps for (a) the best $\boldsymbol{U_{\text{gust}}}$ model and (b) the best $\boldsymbol{Z}$ model in physical units (m s$^{-1}$). The nonlinear, GEV-informed transformation reduces the west-east error gradient in $\boldsymbol{U_{\text{gust}}}$ and improves gust predictions in windstorm-prone Northwestern Europe}
 \label{fig:errormaps}
 \end{figure}

Why do the $\boldsymbol{Z}$ models outperform the $\boldsymbol{U_{\text{gust}}}$ models on the test set while underperforming on training and validation sets? To address this, we examine whether this performance gap varies across geographical regions. We compare the best $\boldsymbol{U_{\text{gust}}}$ and $\boldsymbol{Z}$ models trained with the ``ICP split'' feature selection method to assess regional differences in model skill. The $\boldsymbol{U_{\text{gust}}}$ model exhibits a distinct west-east RMSE gradient (Fig. \ref{fig:errormaps}a), with higher errors in Northwestern Europe, where severe windstorms frequently make landfall. This suggests that the model prioritizes inland clusters with weaker winds, potentially at the expense of distinguishing differences in strong coastal gusts (Fig. \ref{fig:errormaps}a). It may also struggle to anticipate rare inland windstorms, such as Storm \textit{Aila} (2020) \citep{rantanen2021}. Therefore, the main difference between the $\boldsymbol{U_{\text{gust}}}$ and $\boldsymbol{Z}$ error maps (Fig. \ref{fig:errormaps}) is that the nonlinear gust transformation imposes stronger spatial regularization. The $\boldsymbol{Z}$ model shows decreased skill in Eastern Europe and the Balkans but improved performance in Northwestern Europe, suggesting reduced geographical bias. These results indicate that the GEV-informed gust transformation mitigates regional biases in MLR predictions, leading to a fairer model despite its slightly higher overall RMSE compared to $\boldsymbol{U_{\text{gust}}}$ MLRs.

\subsection{Leveraging Interpretable, Pareto-Optimal Equations for Scientific Insight} \label{section:physical}
Both the best 4-feature $\boldsymbol{U_{\mathrm{gust}}}$ and $\boldsymbol{Z}$ models are linear functions of the \textit{standardized} PC loadings, defined as $\widetilde{\mathrm{PC}}_{X,j} = \frac{\mathrm{PC}_{X,j} - \mathrm{PC}_{X,j}^{\mathrm{mean}}}{\mathrm{PC}_{X,j}^{\mathrm{std}}}$, where the mean and standard deviation are calculated over the training set and $j$ indexes the selected features. After fitting, these standardized PC loadings are multiplied by region-specific weights $\boldsymbol{a_X} \in \mathbb{R}^{15\times4}$ and then added to biases $\boldsymbol{b} \in \mathbb{R}^{15}$ for each of the 15 regions of interest. Using the inner product from Eq.~\ref{define:PC}, we express the models as scaled projections of winterstorm fields onto data-driven temporal patterns $\Pi_{X,j}(t)$. The equation for the $\boldsymbol{U_{\mathrm{gust}}}$ model is: 

\vspace{0pt}
\begin{equation} \label{eq:U_rewritten}  
\boldsymbol{U_{\mathrm{gust}}^{U}} = \boldsymbol{\beta^{U}} + \sum_{j=1}^{4} \boldsymbol{\alpha_{X,j}^{U}} \left\langle \Pi_{X,j}(t) \mid X_{j}(t) \right\rangle _{X_{j}}\  \text{where:}\quad  
\boldsymbol{\alpha^U_{X,j}} = \frac{\boldsymbol{a^U_{X,j}}}{\mathrm{PC}_{X,j}^{\mathrm{std}}}, \   
\boldsymbol{\beta^U} = \boldsymbol{b^U} - \sum_{j=1}^{4} \frac{\mathrm{PC}_{X,j}^{\mathrm{mean}}}{\mathrm{PC}_{X,j}^{\mathrm{std}}} \boldsymbol{1}.  
\end{equation}
\vspace{-10pt}
For the $\boldsymbol{Z}$ model, using the definition of $Z$ and the GEV quantile function from Eq.~\ref{eq:GEV}, we obtain:  

\begin{equation} \label{eq:Z_rewritten}  
\boldsymbol{U_{\mathrm{gust}}^{Z}} = \boldsymbol{G}_{\boldsymbol{\mu,\sigma,\xi}}^{-1}  
\left( 1 - \exp \left[ -\boldsymbol{\beta^{Z}} - \sum_{j=1}^{4} \boldsymbol{\alpha_{X,j}^{Z}}  
\left\langle \Pi_{X,j}(t) \mid X_{j}(t) \right\rangle _{X_{j}} \right] \right)\  \text{with}\left(\boldsymbol{\alpha^Z_{X,j}},\boldsymbol{\beta^Z}\right) \text{similarly\ defined}. 
\end{equation}

$\boldsymbol{U_{\mathrm{gust}}^{U}}$ and $\boldsymbol{U_{\mathrm{gust}}^{Z}}$ are the predicted maximum wind gusts for each region using the direct and transformed models, respectively; $\boldsymbol{b^{U}}, \boldsymbol{b^{Z}} \in \mathbb{R}^{15}$ are region-specific biases; $\boldsymbol{a_{X,j}^{U}}, \boldsymbol{a_{X,j}^{Z}} \in \mathbb{R}^{15}$ are region-specific regression weights for the selected PC loadings; $\Pi_{X,j}(t)$ are the data-driven temporal patterns associated with each winterstorm feature; $\langle \Pi_{X,j}(t) \mid X_{j}(t) \rangle_{X_{j}}$ represents the projection of the storm field $X_j(t)$ onto the temporal mode $\Pi_{X,j}(t)$; $\boldsymbol{G}_{\boldsymbol{\mu,\sigma,\xi}}^{-1}$ is the GEV quantile function, mapping transformed values back to physical wind gust units; and $\boldsymbol{1} \in \mathbb{R}^{15}$ is the identity vector (a column vector of ones). 

\begin{figure}%
 \centering
 \includegraphics[width=0.99\textwidth]{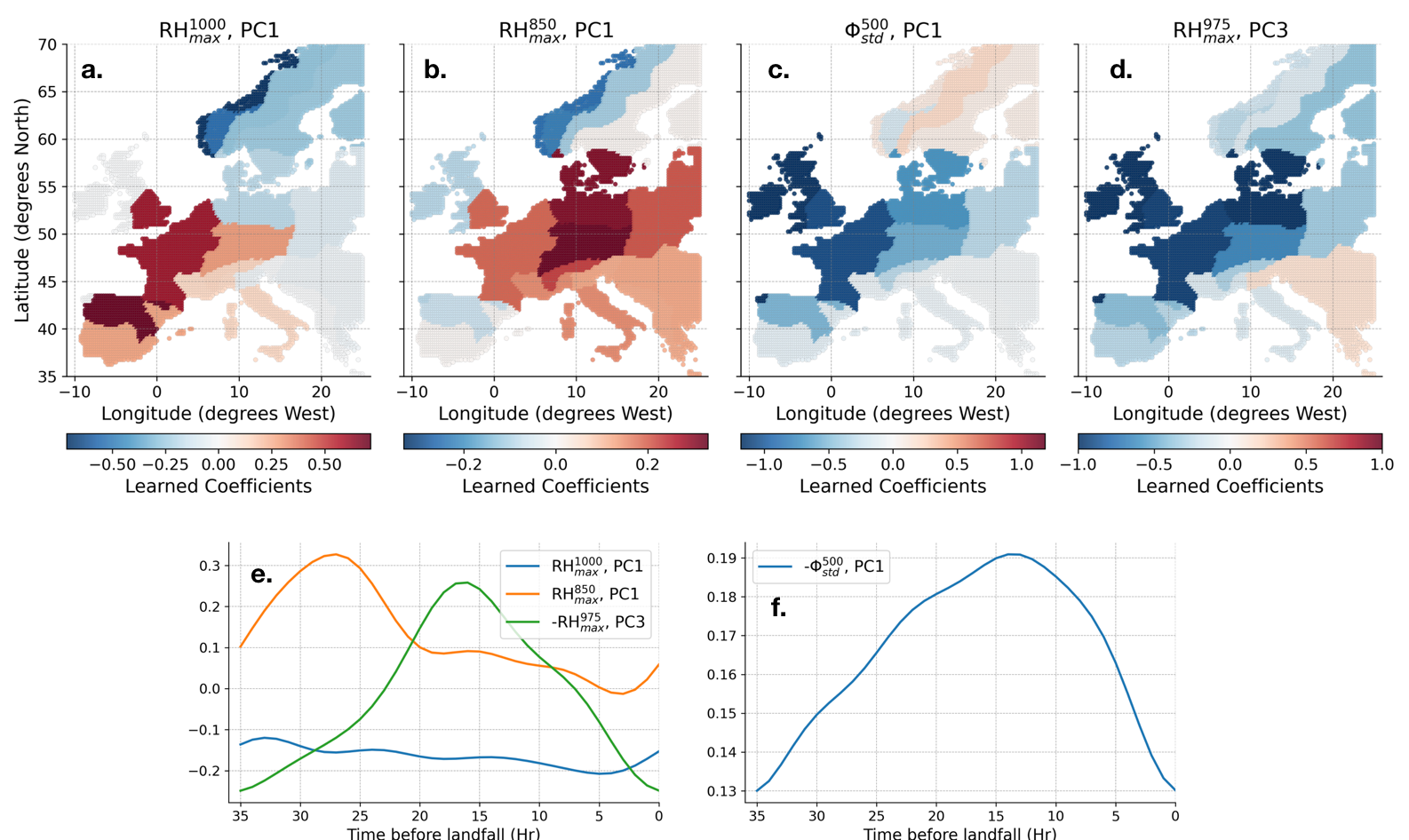}
 \caption{A visual representation of the learned weights for the 4 features in equation of our best $\boldsymbol{Z}$ model: (a) $RH_{\text{max}}^{1000}$, (b) $RH_{\text{max}}^{850}$, (c) $RH_{\text{max}}^{975}$, and (d) $\Phi_{\text{std}}^{500}$. Comparing these weights across clusters to the time series of the four selected PC modes (e-f) highlights key temporal patterns in storm history that promote or suppress extreme winds after landfall in different regions}
 \label{fig:PCcomp}
 \end{figure}

For a region \(i\) and a feature \(j\), models predict higher \(U_{\mathrm{gust},i}\) when the product of \(a_{X,j,i}\) and \(\widetilde{PC}_{X,j}\) is positive, and vice versa. Assuming \(a_{X,j,i} > 0\), time series that project positively onto the PC eigenvectors lead to higher wind gust predictions. The regression coefficients \(\boldsymbol{a_{X,i}}\) in the best \(\boldsymbol{Z}\) model exhibit significant geographical variability (Fig.~\ref{fig:PCcomp}a-d), indicating that the influence of temporal patterns on gusts varies across regions. For example, two low-tropospheric humidity PCs in the \(\boldsymbol{Z}\) model are positively correlated with gusts in region 4 (France and the Netherlands; Fig.~\ref{fig:PCcomp}a-b), while the other two PC terms are negatively correlated (Fig.~\ref{fig:PCcomp}c-d). The temporal evolution of the first PC mode of \(RH_{\text{max}}^{850}\) and the third PC mode of \(RH_{\text{max}}^{975}\) (Fig.~\ref{fig:PCcomp}e) suggests that higher gusts in France occur when drying persists within or at the top of the boundary layer (850 hPa) before storm landfall. A drier lower troposphere promotes the descent of high-speed winds via evaporation \citep{browning2015} or through downward mixing \citep{pantillon2018}. In contrast, a larger standard deviation of 500 hPa geopotential height (\(\Phi\)) before landfall (Fig.~\ref{fig:PCcomp}f) may indicate more perturbed west-east winds, which can intensify winterstorms. To summarize, Fig.~\ref{fig:PCcomp} demonstrates that we can extract important storm predictors leading to extreme wind gusts, such as low-tropospheric humidity, and quantify their relative importance. Finally, the temporal variability encoded in the PCs reveals additional insights into the timing of the drying or moistening tendency relative to landfall, which increases the likelihood of strong gusts.

\section{Conclusion}
We derived interpretable equations linking storm history to post-landfall extreme wind gusts through a PC-regression-based framework. By combining dimensionality reduction and temporal smoothing, we trained a hierarchy of equations from small datasets while mitigating overfitting via careful cross-validation, sequential feature selection, and Pareto optimization. Our results suggest that hierarchical modeling improves generalizability in small-sample settings, with the ``Invariant Causal Prediction split'' providing the greatest benefit for unseen test cases. Pareto optimization indicates that equations should contain at most 3–4 features to balance generalizability and predictive skill. While the best-performing $\boldsymbol{U_{\text{gust}}}$ model performs well globally (mean validation RMSE), spatial analysis reveals a west-east error structure, as the model prioritizes weak-wind regions at the expense of capturing gust variations in strong-wind regions. A GEV-informed nonlinear transformation reduces this spatial bias by accentuating distinctions in gust percentiles at the distribution tail. The discovered equations provide insights into how the timing of low-tropospheric drying is associated with wind gusts in Northwestern Europe, underscoring the importance of storm thermodynamic history in predicting post-landfall wind gusts. Future extensions could apply this framework to larger datasets and explore alternative interpretable modeling approaches, including nonlinear methods such as symbolic regression.

\paragraph{Acknowledgments}
We are grateful for the technical assistance of M. Gomez and the advice of E. Koch.

\paragraph{Funding Statement}
This research was supported by the Canton of Vaud, Switzerland, through the base funding of UNIL.

\paragraph{Competing Interests}
None.

\paragraph{Data Availability Statement}
The released version of the Github repository containing code to reproduce the main figures of this work can be accessed through Zenodo and has been assigned the following DOI: \href{https://doi.org/10.5281/zenodo.14766993}{10.5281/zenodo.14766994}. The ECMWF Windstorm Indicator Data can be downloaded from the Copernicus \href{https://cds.climate.copernicus.eu/datasets/sis-european-wind-storm-indicators?tab=overview}{Windstorm Indicator} data archive. ERA5 datasets were downloaded from the Copernicus website (\href{https://cds.climate.copernicus.eu/cdsapp#!/dataset/reanalysis-era5-pressure-levels?tab=form}{multiple pressure levels} as well as \href{https://cds.climate.copernicus.eu/cdsapp#!/dataset/reanalysis-era5-single-levels?tab=form}{single pressure levels}).

\paragraph{Ethical Standards}
The research meets all ethical guidelines, including adherence to the legal requirements of the study country.

\paragraph{Author Contributions}
All authors contributed to the conceptualization, methodology, writing, and approved the final submitted draft. Data curation and visualization: F.I.T., F.A.

\bibliographystyle{plainnat}
\bibliography{references}  






\end{document}


\maketitle


\section{Introduction}
The document expanded upon several methodological details discussed in the main text. The first section (Section A) justifies the use of a manually-selected K number in the K-means++ clustering algorithm to separate different geographical regions. We will show that an optimal K of 4 is insufficient to separate different European regions. For example, all European mountain ranges like the Alps and the Pyrenees are categorized into one cluster using the optimal number of K. The second section (Section B) provides the fitted GEV coefficients for the geographical clusters and the statistical goodness-of-fit test results to ensure that the wind characteristics follow a GEV distribution. The third section (Section C) provides the mean characteristics of the storm cases used in the ``ICP split'' cross-validation method, which achieves the best generalizable equations in our case. The fourth section (Section D) describes the variables used in the best 4-feature $\boldsymbol{U_{\mathrm{gust}}}$ and $\boldsymbol{Z}$ models and the learned coefficients of the $\boldsymbol{Z}$ equation. The final section (Section E) summarizes the physical variables used to generate the storm history information used in this work.

\section*{A. The optimal number of K clusters is insufficient in finding coherent subregions}
\begin{figure}[htb]
  \centering
  \includegraphics[width=0.7\textwidth]{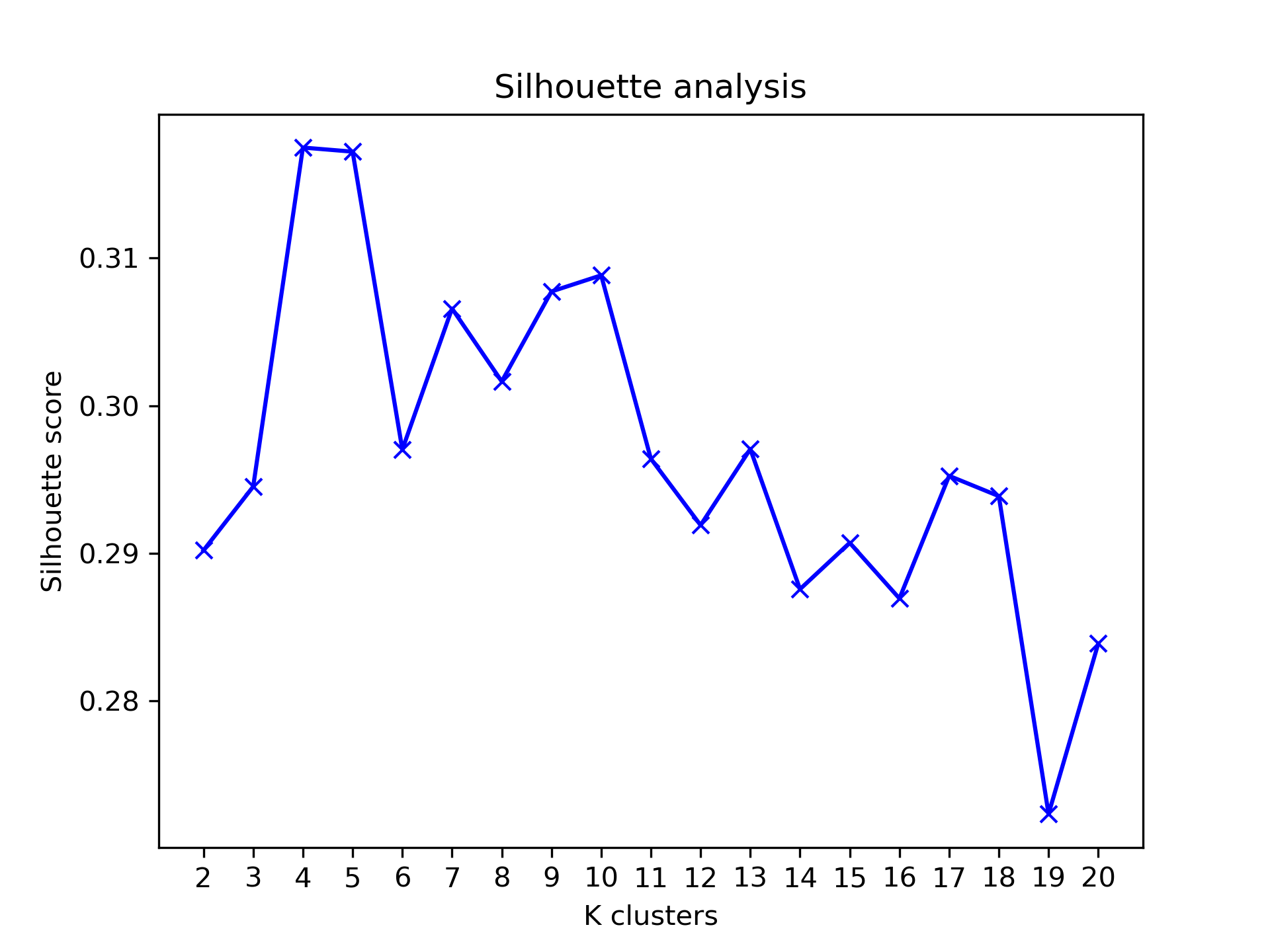}
 \caption{Silhouette analysis of the optimal K number to find coherent geographical subregions}
 \label{SIfig:silhouette}
 \end{figure}

\begin{figure}[htb]
  \centering
  \includegraphics[width=0.7\textwidth]{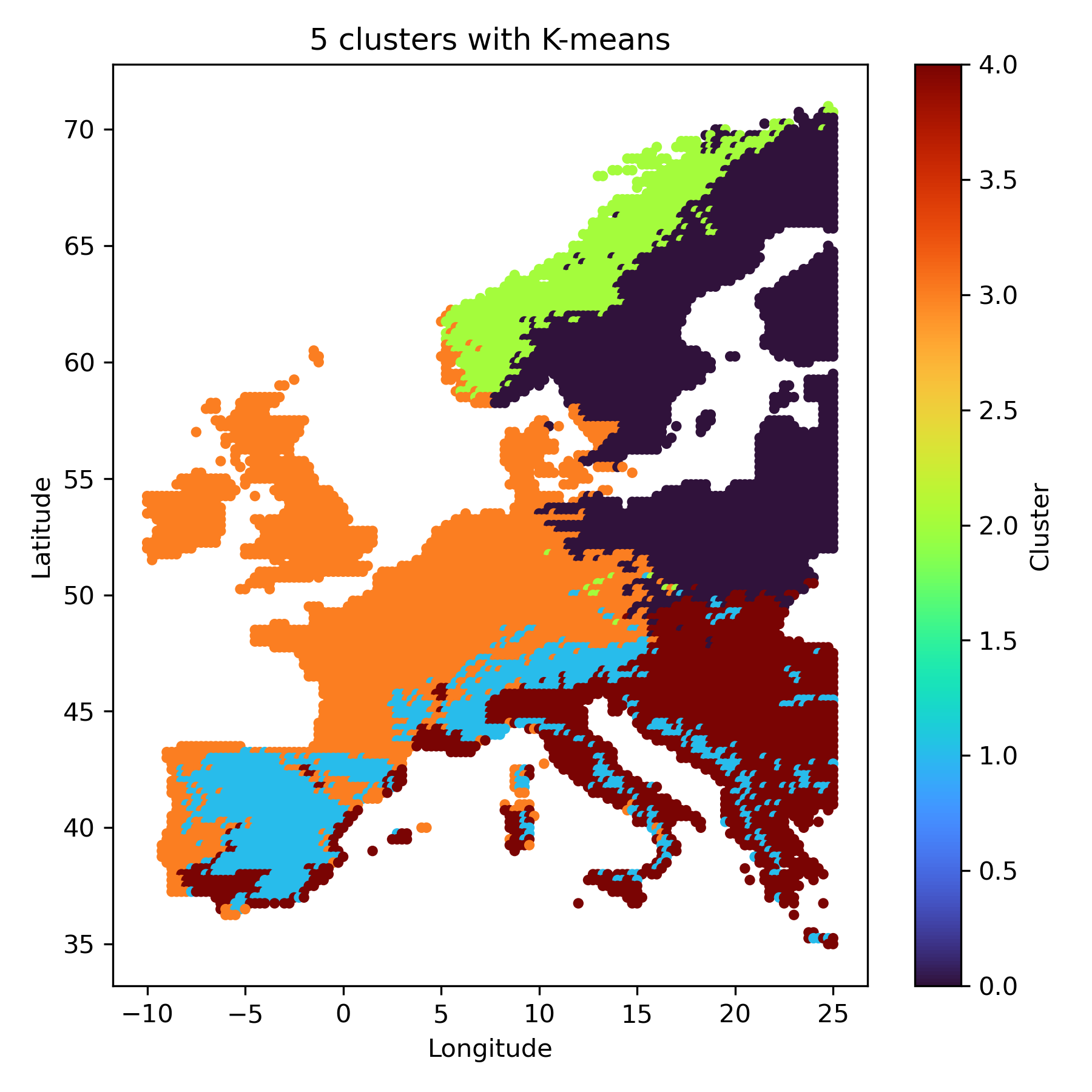}
 \caption{Classification of land grid points using the optimal number of K=5}
 \label{SIfig:clustermap}
 \end{figure}

\begin{figure}%
  \centering
  \includegraphics[width=0.89\textwidth]{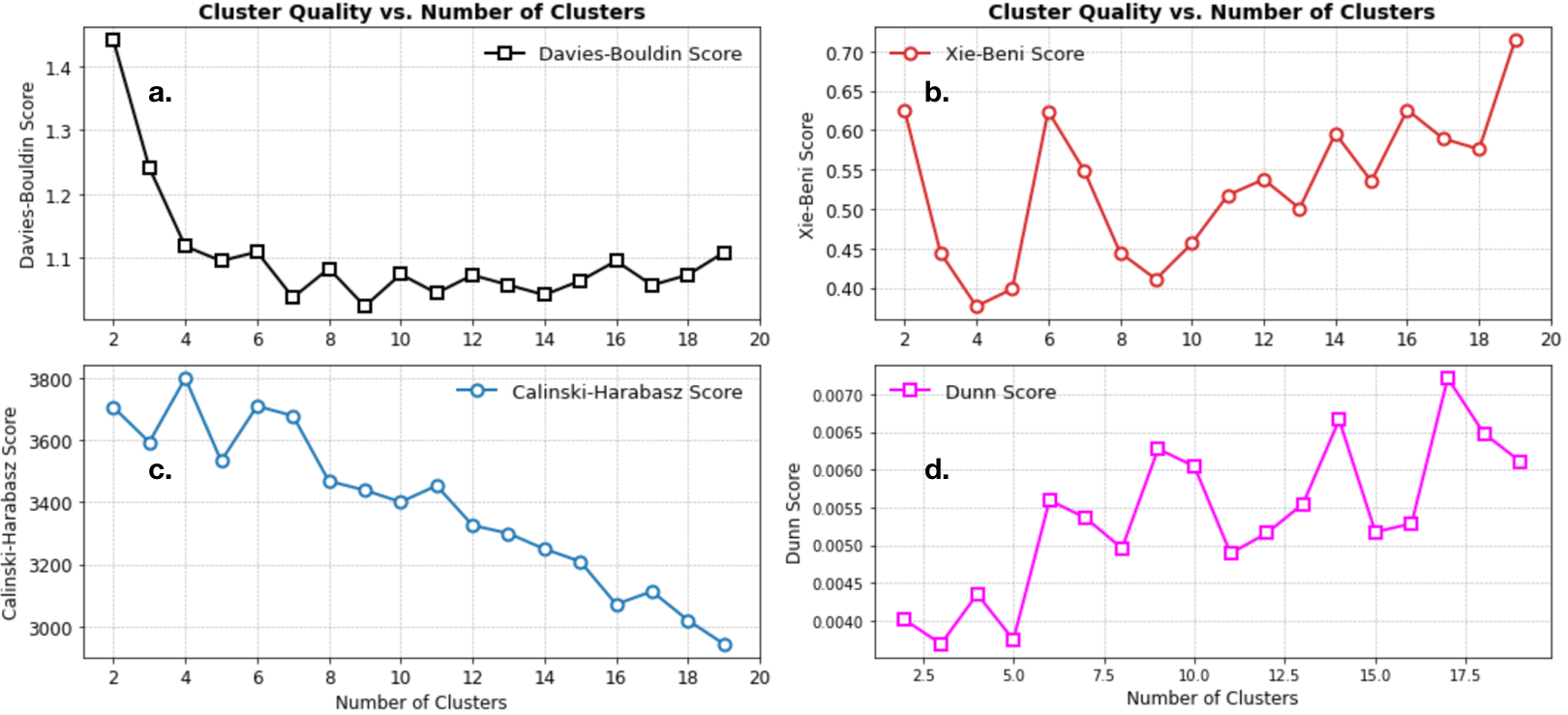}
 \caption{Cluster quality assessments with four different metrics. The upper row are metrics that suggests better cluster quality when the values are low, whereas the lower row shows metrics where higher values mean better cluster quality.}
 \label{fig:response_cluster1}
 \end{figure}

The section justifies the manual selection of 15 clusters during output preprocessing. The optimal cluster number when using a K-means clustering algorithm can be determined with the Silhouette Score analysis. Applying the Silhouette Score to our dataset with gusts conditions, surface elevation, longitude, and longitude yields an optimal K between 4 and 5 (Fig. \ref{SIfig:silhouette}). However, the optimal number of K=5 is insufficient to find coherent geographical clusters when all mountain ranges in Europe are classified as the same cluster, which is too wide for accurate assessment of the gust risk for synoptic-scale extratropical cyclones (Fig. \ref{SIfig:clustermap}).  The other metrics in Figure ~\ref{fig:response_cluster1} do suggest higher optimal cluster numbers. The Davies-Bouldin score suggests cluster numbers higher than 7, with 9 being the optimal number suggested by this metric. While the Xie-Beni score optimizes at a low cluster number of 4, it does have a local minima at 15 clusters. Finally, the Dunn score has a local maxima at the cluster number of 14. 

\begin{figure}%
   \centering
  \includegraphics[width=0.89\textwidth]{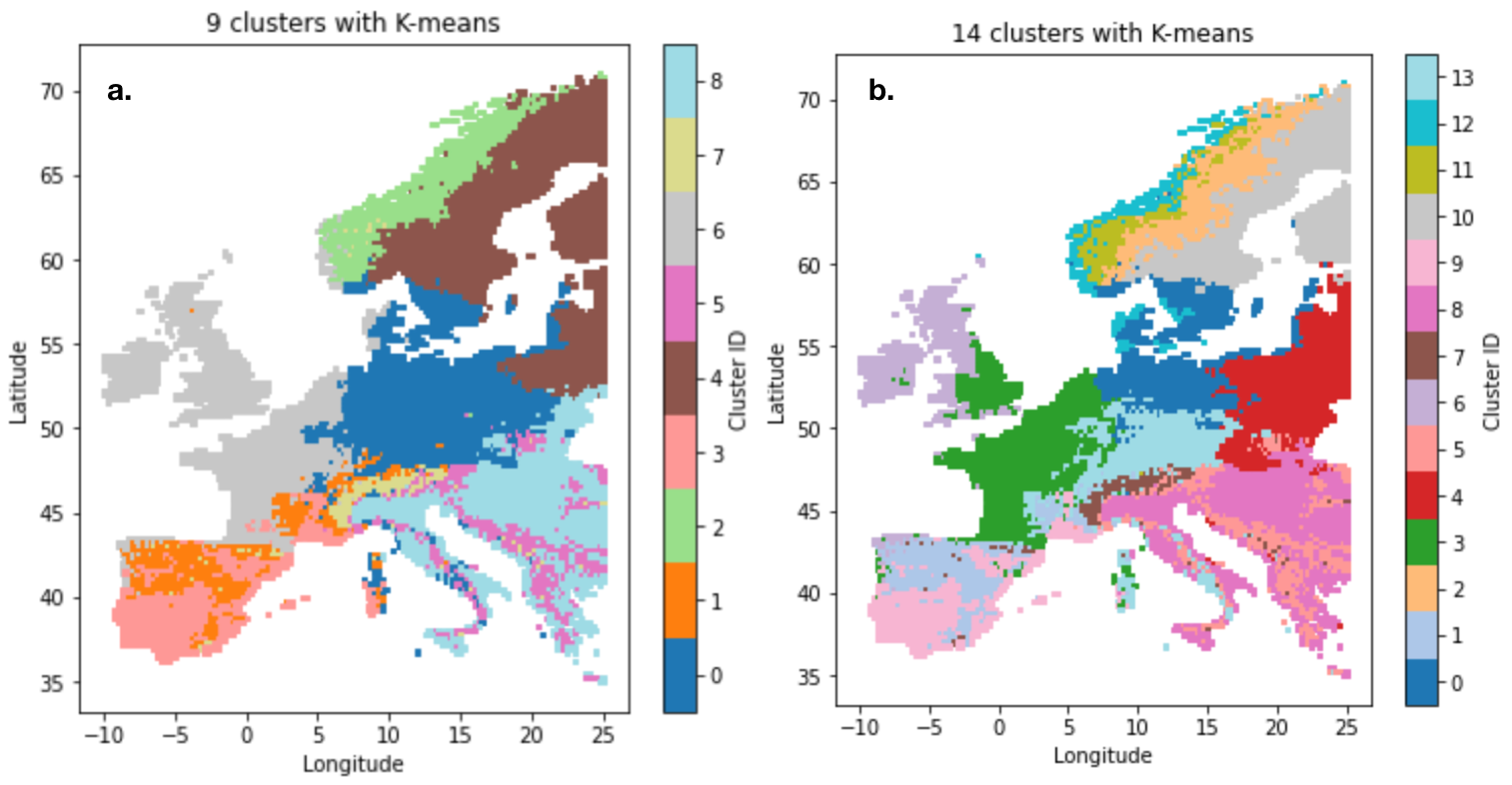}
 \caption{Raw, i.e., not smoothed with kNN algorithm, cluster maps with an optimal cluster number of (a) 9, and (b) 14.}
 \label{fig:response_cluster2}
 \end{figure}

Summarizing these results, we have 3 additional cluster number candidates in addition to the 4 clusters that we rejected on the basis of overly large clusters. These candidates are 9, 14, and 15. Figure ~\ref{fig:response_cluster2} shows the unsmoothed version of the cluster maps with 9 and 14 clusters. Compared to our final cluster map, both options do not achieve the right geographical coherence and regional specificity for our purpose, particularly in southern Europe and the Balkans. Both of these cluster maps consider a large part of Italy and the Carpathian Basin to be a single coherent entity. This is problematic as it is too large and violates established European climate regimes.

\section*{B. GEV distributions} \label{secGEV}
\subsection*{Fitted GEV coefficients for the geographical clusters}
We provide a list of all fitted GEV coefficients for the 15 geographical clusters for preparing the nonlinear transformed $Z_{gust}$ and Figure 1 in the main text.

\begin{table}[htb] \label{Table:GEV}
\begin{tabular}{l|lll}
Cluster number & Location & Scale  & Shape \\ \hline
1              & 0.073    & 16.247 & 4.020 \\
2              & 0.033    & 14.911 & 4.533 \\
3              & 0.032    & 16.805 & 4.429 \\
4              & 0.135    & 18.439 & 4.398 \\
5              & 0.094    & 15.623 & 3.549 \\
6              & 0.138    & 17.691 & 4.486 \\
7              & 0.103    & 21.075 & 5.003 \\
8              & -0.023   & 13.722 & 4.563 \\
9              & 0.132    & 17.204 & 3.982 \\
10             & 0.137    & 16.468 & 4.272 \\
11             & 0.047    & 16.844 & 3.795 \\
12             & 0.120    & 16.232 & 5.364 \\
13             & 0.102    & 19.381 & 4.827 \\
14             & 0.128    & 17.504 & 4.564 \\
15             & 0.112    & 15.716 & 4.879
\end{tabular}
{\caption{The fitted GEV coefficients of the 15 clusters}}
\end{table}

\subsection*{Statistical fitness tests to determine the definition of block maxima}
\begin{figure}
   \centering
  \includegraphics[width=0.97\textwidth]{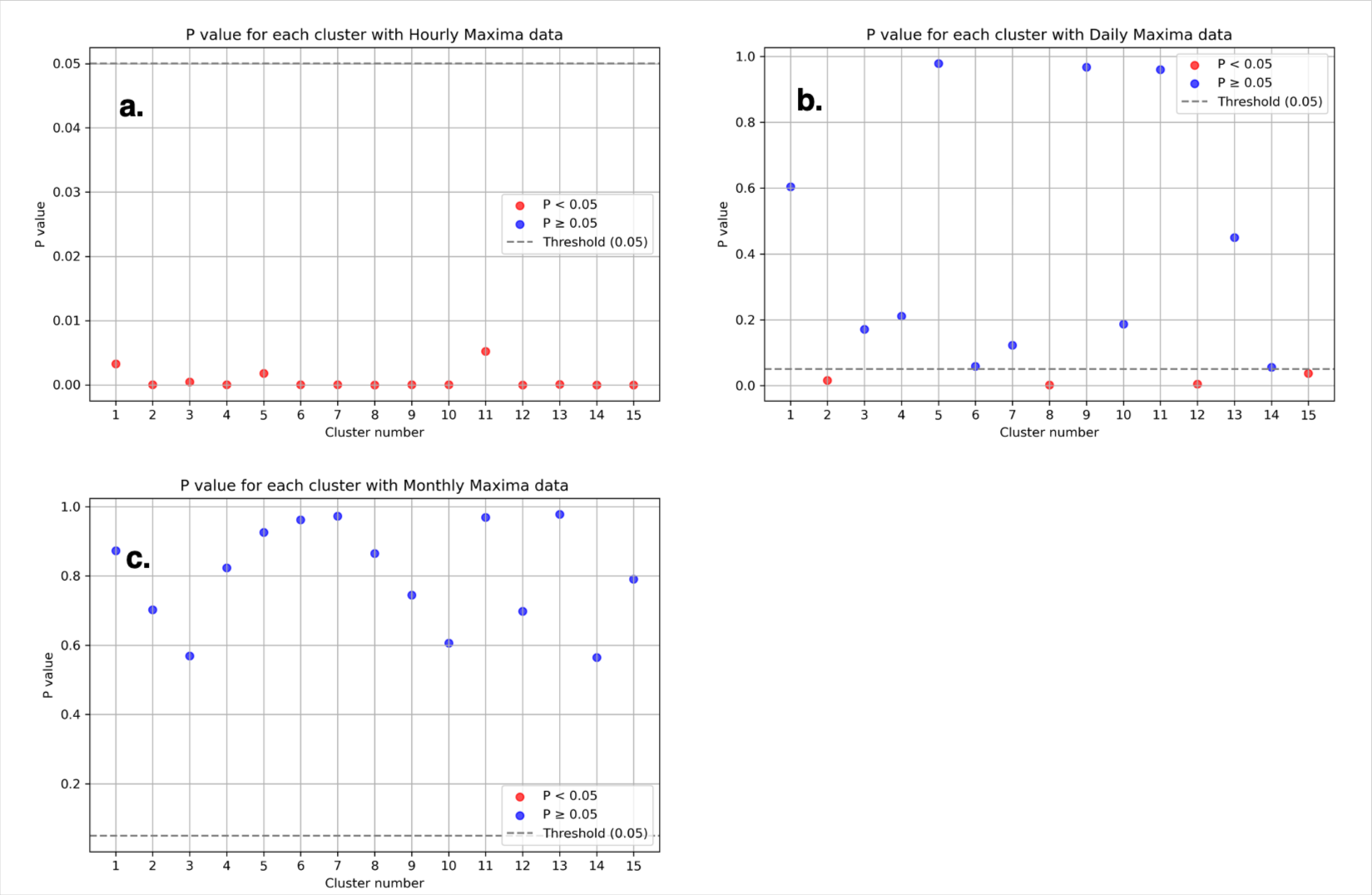}
 \caption{Statistical fitness test for different definitions of block maxima: hourly (a), daily (b), and monthly (c)}
 \label{SIfig:KS}
 \end{figure}

To determine the definition of block maxima suitable for our data, we compare the empirical distribution to the fitted GEV distribution under different definitions of block maxima to test the goodness of fit. Using the p-value of Kolmogorov-Smirnov test, we see spatial variance in the GEV goodness-of-fits and a high sensitivity of the goodness-of-fit to block maxima definitions (Fig. \ref{SIfig:KS}). The overall best candidate for analysis was initially determined to be monthly maxima, as all clusters
have p-values greater than the statistical threshold. However, we found that the maximum 15-hour wind speed for the storm events (maximum values) can occasionally fall outside the range of monthly maxima in some low winds regions as this definition removes too many low gust values. This ultimately lead to difficulties in getting to percentile values for all storms and clusters. With this in mind, we ultimately select daily maxima as our choice of block maxima in our study, with the caveat that the gusts in certain geographical regions (e.g., the Alps) cannot be accurately represented by the GEV fit.

\section*{C. Mean wind characteristics of the 7 Storm Case Clusters}
The section shows the mean wind characteristics and the storm case counts of the seven storm case clusters used to produce the cross-validation folds based on storm gust characteristics. The mean gust winds of the storm clusters are shown, along with their corresponding storm counts (Table \ref{tab:mean_winds}). From the table, we see a nice separation of windstorms with different spatial characteristics in their gust distribution. For example, the cluster with the lowest case count (Cluster 6) contains the two storms that had unusually high values in the 15$^{th}$ geographical cluster. 

\begin{table}
    \centering
    \small
    \begin{tabularx}{\linewidth}{c|c|X}
        \toprule
        \textbf{Cluster} & \textbf{Case Count} & \textbf{Mean Winds (m s$^{-1}$)} \\
        \midrule
        1 & 15 & 25.39, 24.21, 16.32, 30.26, 18.49, 19.58, 37.57, 20.51, \newline 17.75, 16.16, 17.87, 20.17, 20.08, 21.36, 18.90 \\
        2 & 11 & 18.65, 16.12, 17.84, 27.14, 14.86, 13.83, 39.46, 12.32, \newline 14.54, 12.00, 17.59, 21.02, 22.05, 14.40, 18.25 \\
        3 & 6  & 31.74, 17.87, 29.68, 26.90, 23.59, 17.65, 42.15, 14.26, \newline 16.04, 15.41, 26.47, 40.16, 41.85, 16.53, 13.94 \\
        4 & 8  & 17.54, 10.79, 27.21, 17.57, 15.86, 18.21, 31.36,  9.18, \newline 18.10, 14.13, 23.38, 29.21, 33.70, 12.99, 12.88 \\
        5 & 10 & 30.58, 28.84, 21.19, 32.72, 23.49, 20.90, 42.69, 20.91, \newline 20.53, 14.67, 21.56, 27.00, 26.86, 20.06, 17.74 \\
        6 & 2  & 18.55, 27.19, 17.95, 37.09, 18.52, 23.08, 37.49, 26.68, \newline 17.48, 35.11, 18.88, 15.64, 18.14, 28.22, 43.35 \\
        7 & 4  & 19.96, 18.61, 23.05, 31.01, 15.72, 15.09, 34.93, 14.84, \newline 15.90, 17.01, 20.41, 25.23, 26.87, 17.78, 28.75 \\
        \bottomrule
    \end{tabularx}
    \caption{Mean wind speeds for different clusters}
    \label{tab:mean_winds}
\end{table}

\section*{D. The sparse $U_{gust}$ and $Z_{gust}$ models} \label{secSPARSE}
\subsection*{Selected variable list}
This table describes the variables used in the best-performing sparse $U_{gust}$ and $Z_{gust}$ used to create Figure 3 in the main text. $\mathrm{RH}$ in the table represent relative humidity. The numerical information in the relative humidity terms represents the pressure levels of the selected relative humidity variable. The ``max'' in the variable list is the maximum value spatial filter, ``mean'' the mean value filter, and ``std'' the standard deviation filter. The $\overline{SW_{net}}$ term in the table represents net shortwave radiation flux either at the top of the atmosphere (${TOA}$) or at the surface (${surf}$).

\begin{table} \label{Table:Variablelist_bestZmodels}
\begin{tabular}{l|ll|ll}
                   & \multicolumn{2}{c|}{U\_gust}                                                               & \multicolumn{2}{c}{Z\_gust}                                  \\ \hline
Order of selection & \multicolumn{1}{l|}{Variable name}                                          & PC component & \multicolumn{1}{l|}{Variable name}                       & PC component \\ \hline
1                  & \multicolumn{1}{l|}{$\mathrm{RH}_{max}^{1000}$}                          & 1            & \multicolumn{1}{l|}{$\mathrm{RH}_{max}^{1000}$} & 1       \\
2                  & \multicolumn{1}{l|}{$\mathrm{RH}_{max}^{850}$}                           & 1            & \multicolumn{1}{l|}{$\mathrm{RH}_{max}^{850}$}  & 1       \\
3                  & \multicolumn{1}{l|}{$\overline{SW_{net}^{TOA}}_{std}$}      & 3            & \multicolumn{1}{l|}{$\Phi_{std}^{500}$}        & 1       \\
4                  & \multicolumn{1}{l|}{$\overline{SW_{net}^{surf}}_{mean}$} & 3            & \multicolumn{1}{l|}{$\mathrm{RH}_{max}^{975}$}  & 3      
\end{tabular}
{\caption{Variables chosen by the best $U_{gust}$ and $Z_{gust}$ model}}
\end{table}

\subsection*{Learned coefficient for the best $Z_{gust}$ model}
Table \ref{tab:learnedweights} summarizes the learned coefficient of the best $Z_{gust}$ model.

\begin{table}
    \centering
    \small
    \begin{tabular}{c|cccc}
        \toprule
        \textbf{Cluster} & \textbf{$\mathrm{RH}_{max}^{1000}$} & \textbf{$\mathrm{RH}_{max}^{850}$} & \textbf{$\Phi_{std}^{500}$} & \textbf{$\mathrm{RH}_{max}^{975}$} \\
        \midrule
        1  & -0.1853 &  0.3086 & -0.7057 & -1.1983 \\
        2  &  0.3043 &  0.3302 & -0.5751 & -0.6621 \\
        3  & -0.2852 & -0.1124 &  0.2235 & -0.1178 \\
        4  &  0.6117 &  0.1952 & -1.0632 & -0.9501 \\
        5  & -0.0888 &  0.2108 & -0.3144 & -0.3491 \\
        6  & -0.0082 &  0.1228 & -0.0277 &  0.1537 \\
        7  & -0.0086 & -0.0914 & -1.1815 & -1.2832 \\
        8  &  0.1560 &  0.2300 & -0.3184 & -0.3114 \\
        9  & -0.0367 &  0.1280 & -0.0438 &  0.1542 \\
        10 &  0.2830 &  0.0074 & -0.1139 & -0.2732 \\
        11 & -0.2613 &  0.0100 &  0.0746 & -0.3957 \\
        12 & -0.5521 & -0.2310 & -0.2025 & -0.2643 \\
        13 & -0.7420 & -0.2545 &  0.1037 & -0.1978 \\
        14 &  0.1403 &  0.1564 & -0.0766 & -0.1255 \\
        15 &  0.7162 & -0.0719 & -0.5385 & -0.4183 \\
        \bottomrule
    \end{tabular}
    \caption{The mean learned weights values of the best $Z_{gust}$ model for the 15 clusters across the server splits.}
    \label{tab:learnedweights}
\end{table}

\subsection*{Normalization coefficients for the best $Z_{gust}$ model}
Here, we show the normalization coefficients used to standardize the PC time series for the four PC components in the best $Z_{gust}$ model.
\begin{table}[h]
    \centering
    \begin{tabular}{lcc}
        \hline
        \textbf{Variable} & \textbf{$\mu$} & \textbf{$\sigma$} \\
        \hline
        $\mathrm{RH}_{max}^{1000}$ & 0.00031 & 0.05437 \\
        $\mathrm{RH}_{max}^{850}$ & -0.01020 & 0.10174 \\
        $\Phi_{std}^{500}$ & 1.05e-17 & 2.77e-17 \\
        $\mathrm{RH}_{max}^{975}$ & 0.00067 & 0.02530 \\
        \hline
    \end{tabular}
    \caption{Normalization coefficients for the four variables in the best $Z_{gust}$ model}
    \label{tab:variables}
\end{table}

\section*{E. Lists of environmental predictors in the dataset} \label{secVARLIST}
The table describes the 28 environmental predictors used to build the storm history dataset. We categorize different predictors into kinematics, thermodynamics, and other predictors for reference.

\begin{table} \label{Table:Variablelist:Inputs}
\begin{tabular}{l|ll}
\multicolumn{1}{c|}{Predictor}                 & \multicolumn{1}{c}{Category} & \multicolumn{1}{c}{Units} \\ \hline
10m wind speeds                                & Kinematics                   & $m\ s^{-1}$                       \\
100m wind speeds                               & Kinematics                    & $m\ s^{-1}$                       \\
10m wind direction                             & Kinematics                    & degrees                   \\
100m wind direction                            & Kinematics                    & degrees                   \\
2m Temperature                                 & Thermodynamics               & K                         \\
2m Dewpoint Temperature                        & Thermodynamics               & K                         \\
Mean surface net LW radiation flux             & Thermodynamics               & $W\ m^{-2}$    \\
Mean surface net SW radiation flux             & Thermodynamics               & $W\ m^{-2}$    \\
Mean surface latent heat flux                  & Thermodynamics               & $W\ m^{-2}$    \\
Surface latent heat flux                       & Thermodynamics               & $J\ m^{-2}$    \\
Mean surface sensible heat flux                & Thermodynamics               & $W\ m^{-2}$    \\
Surface sensible heat flux                     & Thermodynamics               & $J\ m^{-2}$    \\
Mean TOA net LW radiation flux                 & Thermodynamics               & $W\ m^{-2}$    \\
Mean TOA net SW radiation flux                 & Thermodynamics               & $W\ m^{-2}$    \\
Relative Humidity                              & Thermodynamics               & percent                   \\
Mean vertically-integrated moisture divergence & Thermodynamics               & $mm\ s^{-1}$                      \\
Large-scale snowfall                           & Precipitation                & m                         \\
Large-scale precipitation                      & Precipitation                & m                         \\
Mean large-scale precipitation rate            & Precipitation                & $mm\ s^{-1}$                      \\
Mean total precipitation rate                  & Precipitation                & $mm\ s^{-1}$                       \\
Total precipitation/1hour                      & Precipitation                & m                         \\
Mean sea level pressure                        & Other                        & Pa                        \\
Surface pressure                               & Other                        & Pa                        \\
Convective Available Potential Energy          & Thermodynamics               & $J\ kg^{-1}$                     \\
Geopotential Height                           & Thermodynamics               & $m^2\ s^{-2} $                          \\
High Cloud Cover                               & Other                        & percent                   \\
K-index                                        & Thermodynamics               & K                         \\
Total totals index                              & Thermodynamics               & K                        
\end{tabular}
{\caption{The 28 kinematic, thermodynamic, and other environmental predictors used to produce the input dataset}}
\end{table}



